\documentclass[
reprint,
amsmath,
amssymb,
aps,
pra,
]{revtex4-2}
\usepackage[english]{babel}
\usepackage{graphicx}
\usepackage{dcolumn}
\usepackage{float}
\usepackage{color}
\usepackage{amsmath}
\usepackage{braket}
\usepackage[normalem]{ulem} 
\usepackage{subcaption}
\usepackage{ragged2e}

\definecolor{darkgreen}{rgb}{0,0.5,0}
\definecolor{grey}{rgb}{.6,.6,.6}
\definecolor{darkblue}{rgb}{0,0,0.5}

\begin{document}
\title {Semiconductor Circuits for Quantum Computing with Electronic Wave Packets}

\author {David Pomaranski$^{1,2}$$^{*+}$, Ryo Ito$^{1,3}$, Ngoc Han Tu$^{1}$, Arne Ludwig$^{4}$,  Andreas D. Wieck$^{4}$, Shintaro Takada$^{3,5}$, Nobu-Hisa Kaneko$^{3}$, Seddik Ouacel$^{6}$, Christopher B\"auerle$^{6}$, and Michihisa Yamamoto$^{1,2}$$^{*+}$}

\affiliation{$^{1}$RIKEN Center for Emergent Matter Science (CEMS), Wako-shi, Saitama, Japan\\
$^{2}$Quantum Phase Electronics Center and Department of Applied Physics, The University of Tokyo, Tokyo, Japan\\ 
$^{3}$The National Institute of Advanced Industrial Science and Technology, Tsukuba, Japan\\
$^{4}$Faculty of Physics and Astronomy, Ruhr-University Bochum, Bochum, Germany\\
$^{5}$Department of Physics, Osaka University, Osaka, Japan\\
$^{6}$Institut NEEL, CNRS, France\\
$^{*}$ These authors contributed equally to the work. 
$^{+}$Correspondence should be sent to D.P. (email: david@ap.t.u-tokyo.ac.jp), M.Y. (email: michihisa.yamamoto@riken.jp) }
 
\maketitle

\textbf{Standard approaches to quantum computing require significant overhead to correct for errors. The hardware size for conventional quantum processors in solids often increases linearly with the number of physical qubits, such as for transmon qubits in superconducting circuits \cite{Koch2007,Kjaergaard2020}
or electron spin qubits in quantum dot arrays \cite{Loss1998,Burkard2023}. 
While photonic circuits based on flying qubits do not suffer from decoherence or lack of potential scalability \cite{Knill2001,Takeda2019}
, they have encountered significant challenges to overcome photon loss in long delay circuits. 
Here, we propose an alternative approach that utilizes flying electronic wave packets propagating in solid-state quantum semiconductor circuits. Using a novel time-bin architecture for the electronic wave packets, hardware requirements are drastically reduced because qubits can be created on-demand and manipulated with a common hardware element, unlike the localized approach of wiring each qubit individually. The electronic Coulomb interaction enables reliable coupling and readout of qubits. Improving upon previous devices \cite{yamamotoElectricalControlSolidstate2012}, we realize electronic interference at the level of a single quantized mode that can be used for manipulation of electronic wavepackets. This important landmark lays the foundation for fault-tolerant quantum computing with a compact and scalable architecture based on electron interferometry in semiconductors.}

\begin{center}
    \textbf{INTRODUCTION}
\end{center}

Following recent advances in quantum information processing, the realization of fully error-corrected logical qubits consisting of many physical qubits is recognized as one of the major milestones \cite{DiVincenzo1996,Gottesman1998,Reiher2017}.
Considering the fidelity of quantum operations achieved with modern solid-state systems, it is expected that one logical qubit can require thousands of auxiliary physical qubits. Such a fault-tolerant quantum computer would require millions of physical qubits with independent control hardware, creating significant challenges for implementation on a dilution refrigerator that can accommodate only several thousand cables \cite{Krinner2019,Lecocq2021}. 
Quantum computers based on photon flying qubits are being developed with completely different strategies. As an example, reference \cite{takedaUniversalQuantumComputing2017} suggests that universal operations can be performed in an optical loop circuit with only a few embedded fundamental operation (gate) circuits. Losses and difficulties with deterministic control and detection of a single photon still remain a major challenge for universal quantum computing with photons.

\begin{center}
    \textbf{ARCHITECTURE}
\end{center}

While electron coherence in one-dimensional electron systems at the lowest achievable temperature is primarily limited by electron-electron scattering occurring on order of several 100 ps, a wave packet excited by an electronic pulse comparable with or shorter than this can achieve a macroscopic coherence length. 
This is because an electron wave packet (EWP) is the plasmonic charge distribution of an interacting one-dimensional system, known as the eigen-excitation-mode of the Tomonaga-Luttinger liquid \cite{Hashisaka2018, giarmachi2004}.
Such a quasi-particle does not suffer from decoherence induced by electron-electron interactions, as these are already included in the eigenmode of the quasi-particle with the propagation velocity as its eigenvalue. The concept of mean free path does not apply as it does for bare electrons, and the EWP is not back-scattered if there are enough electrons in the channel. Due to the linear dispersion of plasmonic modes arising from Coulomb interactions at low energy, EWPs can propagate without significant distortion \cite{rousselyUnveilingBosonicNature2018}.

Coherence of the EWP can be further extended by applying a Lorentzian-shaped time-dependent voltage pulse to generate a collective excitation that has an integer multiple of the electron charge \cite{Ferraro2014}. 
This quasiparticle (EWP) is free of hole excitations that would normally arise when electrons are excited above the Fermi level. It has an exponential distribution of states in the energy domain only above the Fermi level \cite{levitovElectronCountingStatistics1996a, bauerleCoherentControlSingle2018, glattliLevitonsElectronQuantum2017}. 
Owing to the absence of hole excitations, the energy relaxation induced by coupling to the external environment is also suppressed. High purity electronic excitations with less than 1 $\%$ hole excitations have recently been realized with Lorentzian pulses of temporal width of 30 ps by the frequency comb technique.\cite{aluffiUltrashortElectronWave2023} 

\begin{figure}[!ht]
    \includegraphics[width=0.49\textwidth]{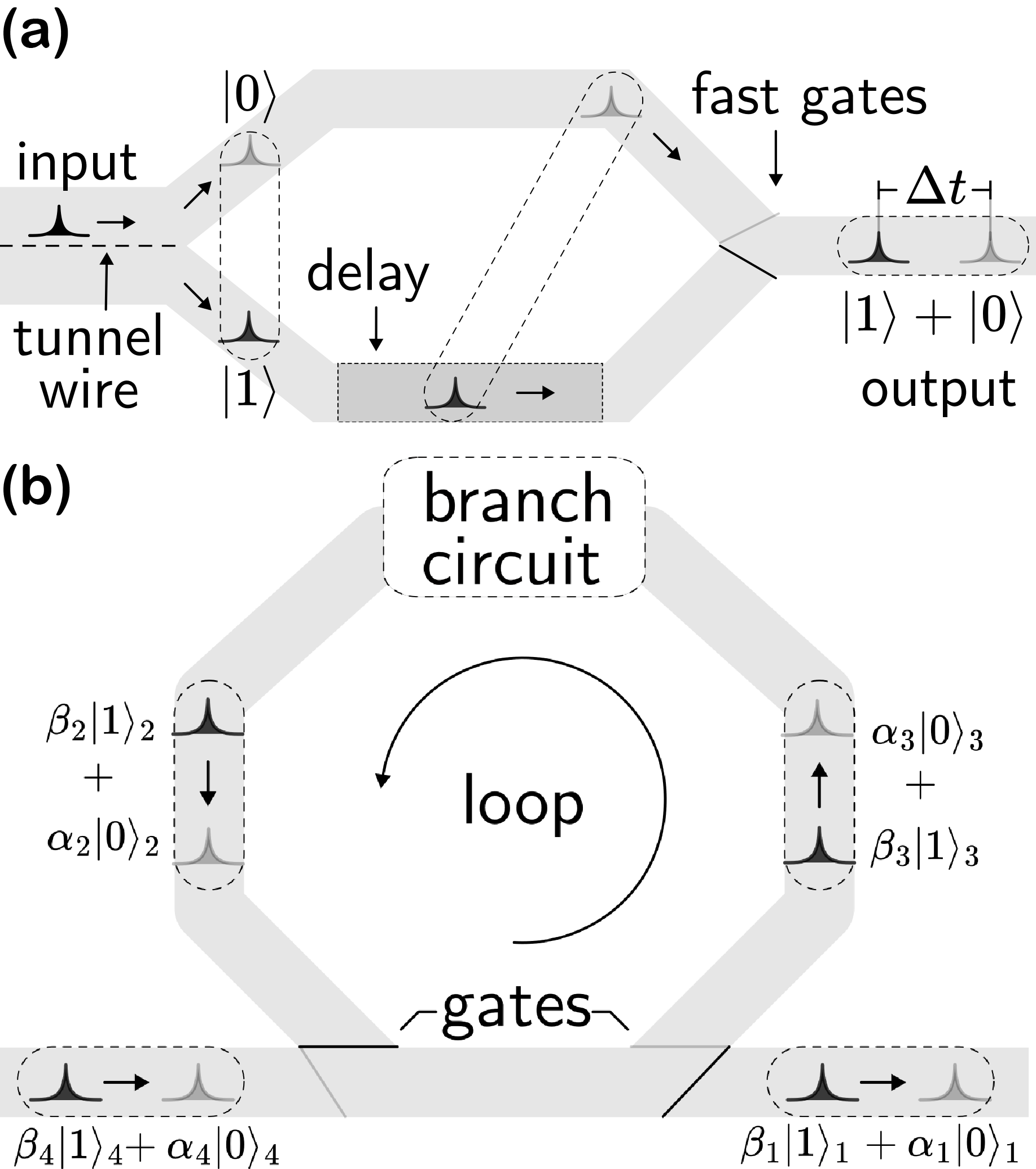}
    \caption{\justifying Illustration of electrical circuits used to generate and manipulate time-bin qubits 
    (a) Creation of a two-path qubit consisting of a superposition state (enclosed by the bubble) in the upper and lower arms, followed by time-bin encoding of the electron wave packets. A time delay $\Delta t$ between the earlier pulse $\ket{0}$ and the later pulse $\ket{1}$ is induced electronically in one arm of the Mach-Zehnder interferometer. The optional fast gate may be used to fully suppress scattering into the other arm of the interferometer. (b) Schematic of a loop circuit that accommodates numerous flying qubits. Time-binned qubits generated by periodic electrical pulses in (a) are guided into the loop circuit where they can branch off into embedded circuits for quantum manipulation. Each flying qubit can traverse the loop multiple times until it is finally ejected for readout, or to be used in a quantum error correction scheme. The intensity of each pulse is used to illustrate the quantum mechanical probability amplitudes of the basis states $|\alpha_i|^2$ and $|\beta_i|^2$.}
    \label{fig:1}
\end{figure}

With these high-purity plasmonic quasiparticles, the qubit is defined as the quantum state of an electronic wave packet (EWP). Conventionally, the two-path qubit has been defined by the presence of an EWP in either path of a two-path interferometer. The most basic two-path interferometer is made up of two parallel, tunnel-coupled quantum wires. However, over long distances, the two-path qubit experiences dephasing during the propagation of the EWP on the dual-rail, caused by the slowly changing electromagnetic environment between the two paths. In order to create a scalable architecture, we propose time-bin encoding for the EWPs. The time-bin qubit is generated using a Mach-Zehnder interferometer with different effective path lengths, as shown in Figure \ref{fig:1}a. In this setup, a delay circuit is introduced into the lower path, which is designed to be longer than the upper path. The delay is further adjusted by modulating the velocity of the propagating plasmonic wave packets, achieved by electrically tuning the width of the path by over an order of magnitude \cite{rousselyUnveilingBosonicNature2018}. A switch gate on the right side of the interferometer directs both the wave packet from the upper path, arriving earlier, and the one from the lower path, arriving later, into a single waveguide. The resulting time-bin qubit is then defined as a superposition of the ``earlier pulse'' ($\ket{0}$) and ``later pulse'' ($\ket{1}$). 
The time-bin configuration has an advantage over the two-path framework because it experiences dephasing only for rapid changes in the circuit potential compared to the separation time between earlier, and later pulses. 
It’s important to note that EWPs are the fastest carriers in semiconductors, making rapid changes in the circuit potential unlikely. Both pulses travel along the same path, encountering almost identical noise conditions, which affect only the global phase of the wavefunction while leaving the relative phase unchanged. For instance, in clean GaAs/AlGaAs 2DEGs, electron-phonon scattering rates ($\tau_{e-ph}^{-1} \approx 1$ ns at 1K) typically scale with $T_{e}^3$ as a function of electron temperature, making these effects negligible over timescales longer than 1$\mu$s at 100 mK \cite{mittalElectronphononScatteringRates1996b}.

By employing time-bin pulses with a 300 ps separation, it should be possible to distinguish between the qubit states $\ket{0}$ and $\ket{1}$ with an efficiency greater than 99$\%$, corresponding to less than 1$\%$ overlap between two Lorentzian EWPs with a temporal width of 30 ps. Although the typical width of such a wave packet, generated using current microwave technology, is limited to around 30 ps, corresponding to a length scale of approximately 10 $\mu$m, sub-picosecond timescales can be achieved through on-chip photoconductive switching and the use of femtosecond optical pulses.\cite{Auston1975,Georgiou2020,bauerleCoherentControlSingle2018}

The train of time-bin qubits created by applying periodic electrical pulses to the interferometer depicted in Figure \ref{fig:1}a is guided into the loop-structure illustrated in Figure \ref{fig:1}b, which accommodates a variety of operation circuits for qubit manipulation (branch circuits) and readout. 
The minimum requirement for universal operation of all the qubits are single-qubit rotation gates, a two-qubit gate, and one measurement (readout) gate. 
Universal operation is performed when the qubits propagate in the loop multiple times while we control the operation circuits with rapid electrical signals.   
This allows for a drastic reduction of hardware size compared to the localized approach where each qubit requires an independent hardware element. The total number of physical qubits is then limited by the number of EWPs in the loop, which should have a length much greater than the coherence length of the time-bin qubit. For a loop of length 100 mm where each qubit occupies 100 $\mu$m, a single loop could accommodate 1000 flying qubits. 

It takes roughly a few hundred ns for the EWPs to circle the loop. While the coherence length has not yet been determined experimentally, measurements using non-Lorentzian excitation with a Mach-Zehnder interferometer in the quantum Hall regime have already demonstrated coherence exceeding 2.5 mm.\cite{hiyamaEdgemagnetoplasmonMachZehnderInterferometer2015} 
With the use of time-bin encoding with Lorentzian-shaped EWPs and suppression of hole excitations by $>$99$\%$, it is expected that the coherence could further increase by several orders of magnitude.

\begin{figure*}[!ht]
    \includegraphics[width=1.0\textwidth]{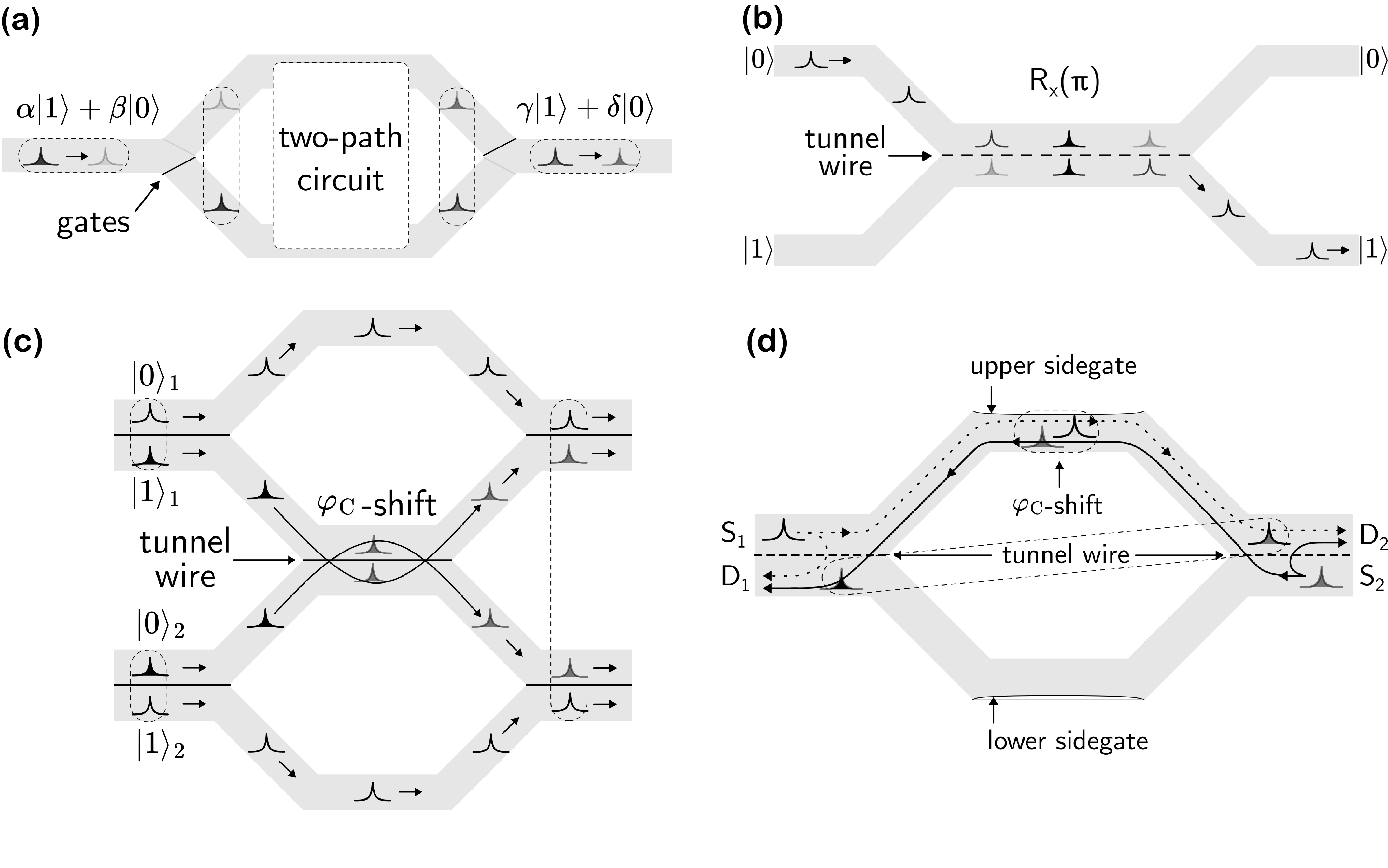}
    \caption{\justifying Schematic illustrations of various qubit manipulation circuits.
    (a) Conversion between the time-bin qubit and the two-path qubit using fast gate branches.
    Gates on the left are rapidly switched to guide the earlier pulse into the upper channel and the later pulse into the lower channel. This can be accomplished with standard microwave electronics on the picosecond timescale. A delay is introduced in the upper path such that both pulses interact within the two-path circuit simultaneously to perform qubit manipulations. After manipulation, another delay is introduced in the lower path, and fast gates on the right to convert the two-path qubit back to the time-bin qubit. 
    (b) Schematic of single-qubit operation with a tunnel-coupled wire demonstrating the implementation of an R$_x(\pi)$ gate. (c) Schematic of a controlled phase gate for two qubits in the two-path representation, corresponding to the operation $\lvert 0_1 0_2 \rangle + \lvert 0_1 1_2 \rangle + e^{i \varphi_C} \lvert 1_1 0_2 \rangle + \lvert 1_1 1_2 \rangle$. Each qubit is prepared as a superposition in the two-path representation (i.e., using a Hadamard gate, or 50-50 beamsplitter). Only when two electrons simultaneously traverse neighboring paths that are coupled by a tunable barrier do they experience the mutual Coulomb interaction, resulting in a phase shift that can also generate an entangled state. The energetic barrier should be tuned such that each EWP returns to its original path after traversing the coupling region. (d) $\varphi_{\rm C}$-shift corresponding to the operation $e^{i \varphi_C} \lvert 0_1 0_2 \rangle + \lvert 0_1 1_2 \rangle + \lvert 1_1 0_2 \rangle + e^{i \varphi_C} \lvert 1_1 1_2 \rangle$. Schematic illustrates the path taken by two interacting electrons propagating simultaneously from S$_1$ and S$_2$. Two EWPs propagating in the upper path are depicted, but they traverse the lower path with equal probability amplitude. Upper/lower side gates can be used to tune the coupling strength between EWPs and control the phase shift. }
    \label{fig:2}
\end{figure*}

Manipulating the time-bin qubit might become challenging due to the short time scale between consecutive states, so we convert it back to the two-path qubit as demonstrated in Figure \ref{fig:2}a to perform gate operations. 
Quantum operations on a two-path flying qubit are implemented using a variety of techniques derived from previous studies of two-path interferometers \cite{yamamotoElectricalControlSolidstate2012,bauerleCoherentControlSingle2018}.  
Single qubit rotation about the x-axis of the Bloch sphere is accomplished by tunnel-coupling two parallel wires (see Figure \ref{fig:2}b). 
Rotation about the z-axis is achieved by adjusting the relative phase between the two-paths, which can be done by electrostatic gating (detuning). This can also be accomplished with an Aharonov-Bohm phase in the presence of a perpendicular magnetic field. These rotation gates allow for universal single qubit operations. 

Two-qubit gates are required to entangle and control multiple qubits, which are achieved by exploiting charge interactions between EWPs. This is a major advantage of our electronic flying qubit over similar photonic architectures. Figure \ref{fig:2}c shows a method for inducing a controlled phase shift, $\varphi_{\rm{C}}$, between two neighboring qubits only when the upper qubit is in the $\ket{1}$ state, and the lower qubit is in the $\ket{0}$ state. Two interferometers are coupled by a tunnel-coupled wire with a tunable energetic barrier, where two adjacent EWPs experience a mutual Coulomb interaction.
If the barrier between neighboring channels is low, Coulomb coupling is also accompanied by undesirable tunnel-coupling that may create a superposition state. Another undesirable effect arises when the barrier is high and using plasmonic EWPs due to charge fractionalization, whereby the propagating eigenmodes generate a charge distribution in adjacent Coulomb-coupled multi-electron transmitting channels \cite{Kamata2014} 
These effects can be suppressed by adjusting the tunnel coupling between adjacent wires and synchronizing the arrival time such that each EWP returns to its original channel after traversing through the coupling region as depicted in Figure \ref{fig:2}c. 
An alternate method to generate entanglement utilizes two counter-propagating EWPs in a single electronic Mach-Zehnder interferometer, as depicted in Figure \ref{fig:2}d. When both EWPs traverse the same arm of the interferometer, they experience the mutual Coulomb interaction resulting in phase shift $\varphi_{\rm{C}}$. When $\varphi_{\rm{C}} = n\pi$, these qubits remain independent, but when $\varphi_{\rm{C}} = \pi/2 + n\pi$ for $n=0,1,2,...$, they become fully entangled \cite{supplementary}.

Another necessary protocol required to implement quantum information processing is the ability to perform single-qubit readout, which can be realized by detecting the presence of a single EWP in a channel. 
One way to accomplish this is by integrating a single-triplet spin qubit in a double quantum dot (S-T qubit) adjacent to a channel, which acts as a pulse detector. 
The propagation of an EWP modifies the exchange interaction of the double quantum dot for a short period of time, which changes the time evolution of the spin qubit state. Based on previous studies, a single EWP consisting of a few electrons can be detected using current technology \cite{thineyInflightDetectionFew2022}. 
Detection sensitivity can be improved by adding a loop circuit such that the EWP makes several traversals in this region. These can be controlled with electrical rather than magnetic means, and have the potential to operate at higher temperatures.

\begin{center}
    \textbf{SINGLE-MODE INTERFERENCE}    
\end{center}
We present here an experimental evaluation of our newly developed quantum operation circuit for two-path qubits in the DC regime. Although this demonstration does not yet involve short EWPs, we demonstrate the feasibility of our two-path interferometer design, as it shows interference in the single transmission mode regime and strong electron-electron interactions—both of which are crucial for implementing the two-qubit gate.

\begin{figure*}[!ht]
    \includegraphics[width=1.0\textwidth]{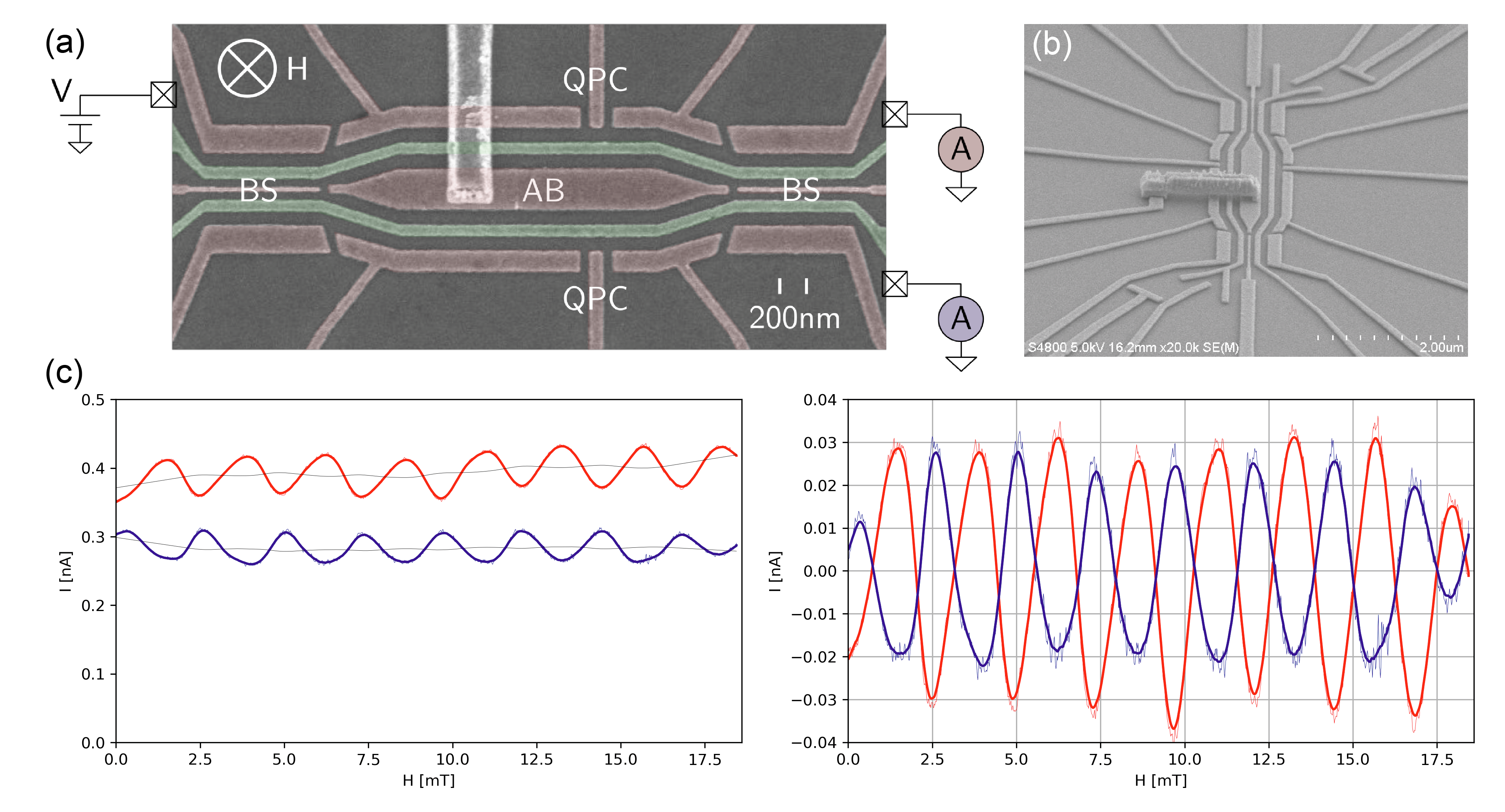}
    \caption{\justifying (a) Scanning electron microscopy image of the flying qubit device with a schematic of the experimental setup. Schottky gates on the surface of an n-AlGaAs/GaAs 2DEG-based heterostructure device (2DEG: $n_D = 3.0\times 10^{11}$ cm$^{-2}$, $\mu \approx 0.88\times 10^{6}$ cm$^2$/Vs, depth = 100 nm) define the device while middle/top gates (highlighted in green) are biased with a positive voltage to increase the localized electron density and screen electrons from impurity potentials. Upper and lower paths are tunnel-coupled in the beam splitter (BS) region by a narrow gate. (b) Device displayed from an angled perspective, highlighting the air bridge structure. (c, left) Currents measured at the BS output oscillate out-of-phase as the relative phase of the electronic wavefunction traversing the central AB region is modified by the weak magnetic field (H) applied perpendicular to the surface.  (c, right) Currents measured after subtracting the steady background, revealing the oscillating components. Thin and thick lines respectively indicate raw and smoothed data.}
    \label{fig:3}
\end{figure*}

The device shown in Figure \ref{fig:3}a is the electronic analog of an optical Mach-Zehnder interferometer that demonstrates coherent phase manipulation of our flying qubit. 
Details of its fabrication and operation can be found elsewhere \cite{yamamotoElectricalControlSolidstate2012, bautzeTheoreticalNumericalExperimental2014, bauerleCoherentControlSingle2018}. 
In an ideal interferometer, the paths should consist of one-dimensional quantum wires to enable phase-coherent transport. However, in real materials, the ballistic length is usually reduced to less than $1 \mu$m when the number of transmitting modes decreases because there is less screening of disorder. Widening the channel allows for multiple transmission modes, increasing the local electron density and helping to screen mobile charge carriers from electrostatic inhomogeneities. 
This also means that the interference signal becomes a complex sum of contributions from all the modes that will cancel each other out. The ideal interferometer should consist of a single, fully transmitting mode that still allows for screening of the disorder potential. To meet these conflicting requirements, narrow metallic Schottky gates were added along the electron trajectory (shown in green color in Figure \ref{fig:3}a). These gates are effective at zero voltage bias due to the screening effect of image charges, which smooth out electrostatic potential fluctuations of charge impurities in the substrate. They also enable the formation of sharper confinement potentials in the quantum wires when the positive voltages are applied to them. Quantum point contacts were also added along the length of each path to allow for quantized mode selection.

While interferometers operating at the level of a single quantized mode of conductance have previously been realized with quantum Hall edge channels \cite{Ji2003},
those devices are not suitable for a scalable architecture based on flying qubits. This is because electrons counter-propagate along chiral edge channels at opposite edges, which requires an ohmic contact at the center of the device. This metallic contact at the center does not preserve the phase-coherence and restricts the scalability. 
While our devices do not encounter fundamental challenges for scalability, they have not been able to operate in the single channel limit due to problems associated with scattering by impurities.

We evaluated the full conductance matrix between four terminals of the interferometer to confirm that the paths correspond to a single mode of conduction, while preserving a high interferometer visibility (see supplemental materials). Visibility is defined as the ratio of amplitude of the interference pattern to the sum of the powers of the components, and it provides a measure of preserved phase coherence. The visibility obtained with the device shown in Figure \ref{fig:3}c is 8.5\%, which is 2 times higher than what was observed in previous designs where multiple modes contribute to the interference \cite{yamamotoElectricalControlSolidstate2012}. 

An important distinction is that those devices could not exhibit finite visibility when transport was limited to a single channel, emphasizing importance of the screening gates.
While the visibility is still far from enabling high fidelity qubit manipulation, a significant improvement in performance is expected upon combining EWPs with the new interferometer design \cite{ouacelElectronicInterferometryUltrashort2024}. 
Major reasons for low visibility observed with DC currents are decoherence induced by the electron-electron interaction, and unwanted back-scattering at the entrance and exit of the ring leading to multiple path contributions. Both of these are suppressed with the use of EWPs that propagate as a plasmonic mode driven by the Coulomb interaction \cite{rousselyUnveilingBosonicNature2018}.
It is expected that combining the pulse excitation with this new interferometer design will further enhance the visibility.

\begin{center}
    \textbf{TWO-ELECTRON INTERFERENCE}    
\end{center}
Recent studies have shown that two propagating electrons repel each other in electronic circuits \cite{Fletcher2023, Ubbelohde2023, wang2023}. However, previous experiments have not yet demonstrated how the Coulomb interaction modulates the interference between the two electrons.

\begin{figure}[!ht]
    \begin{subfigure}[b]{0.49\textwidth}
        \includegraphics[width=\textwidth]{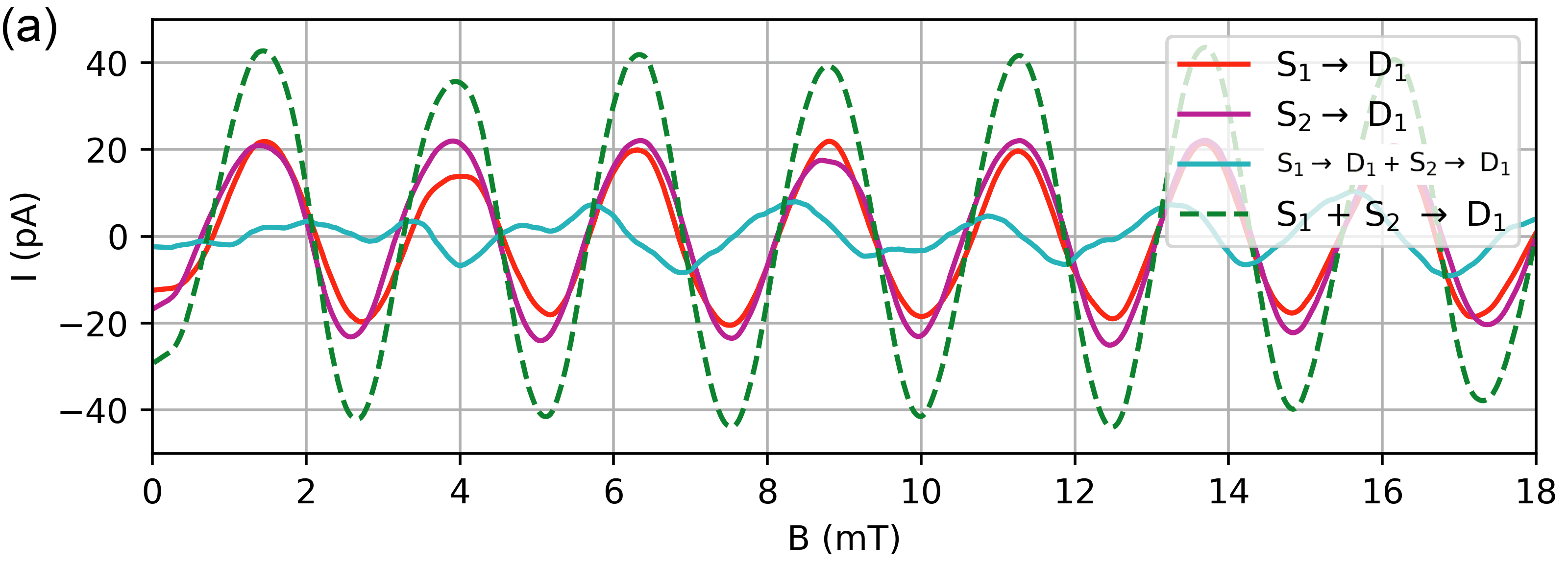}
        \label{fig:4a}
    \end{subfigure}
    \vfill
    \begin{subfigure}[b]{0.49\textwidth}
        \includegraphics[width=\textwidth]{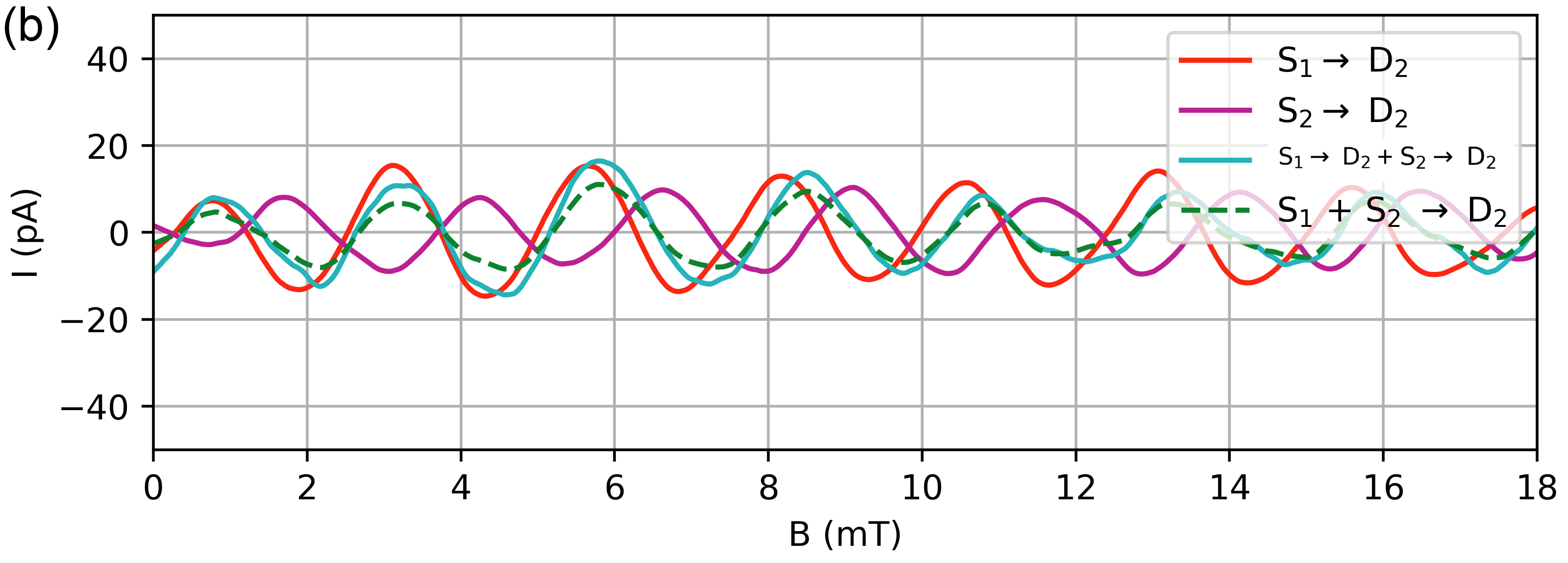}
        \label{fig:4b}
    \end{subfigure}
    \vfill
    \begin{subfigure}[b]{0.55\textwidth}
        \includegraphics[width=\textwidth]{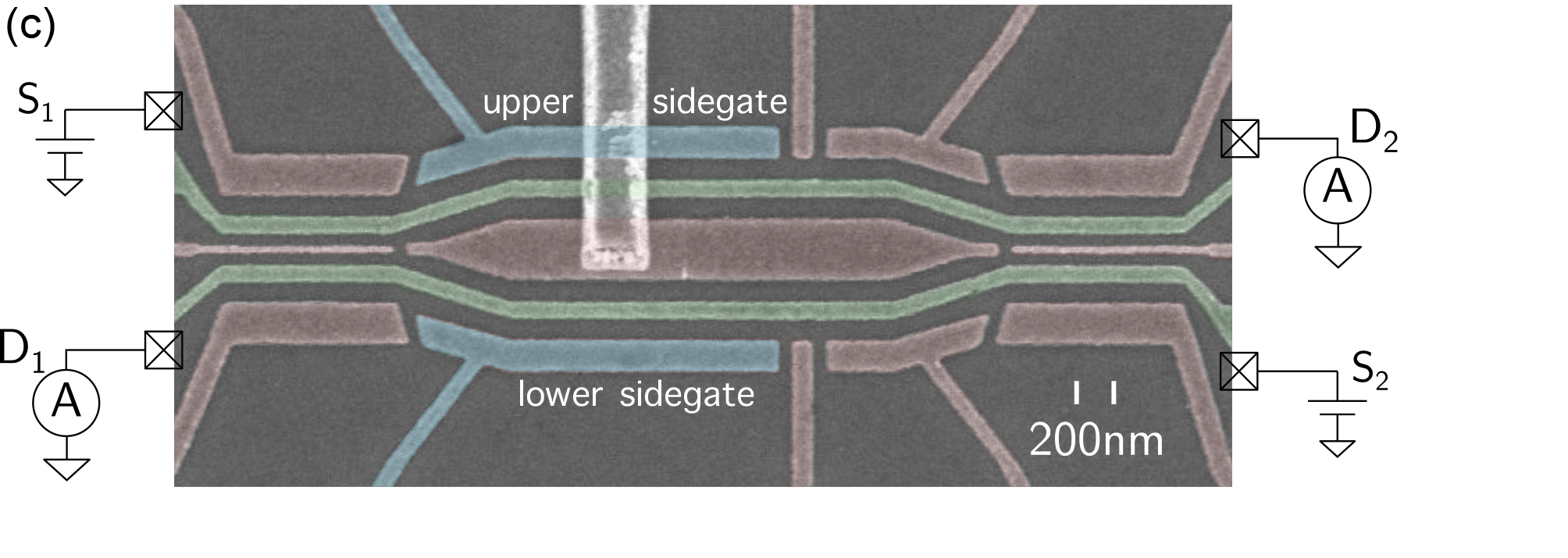}
        \label{fig:4c}
    \end{subfigure}
    \caption{\justifying Experimental data of interference oscillations at D1 when biasing S$_1$ and S$_2$ individually and then simultaneously. In (a), the measured current does not correspond to the sum of the currents S$_1$$\rightarrow$D$_1$ and S$_2$$\rightarrow$D$_1$ when S$_1$ and S$_2$ are biased individually, indicating an additional phase shift which suppresses the interference. In (b), after adjusting the side gate voltages by $\sim 1$ mV, measured current matches the sum of S$_1$$\rightarrow$D$_1$ and S$_2$$\rightarrow$D$_1$, suggesting that $\varphi_{\rm C}$ is a multiple of 0, or $2\pi$. The upper and lower side gate voltages were set to -400 mV and -400 mV in (a), and adjusted to -404 mV and -408 mV in (b).}
    \label{fig:4}
\end{figure}

Two-qubit operations that utilize the mutual Coulomb interaction between propagating EWPs in adjacent Mach-Zehnder interferometers can achieve the controlled phase shift illustrated in Figure \ref{fig:2}c. As an alternative, we implemented a method using counter-propagating charges in a single interferometer (Figure \ref{fig:2}d), by simultaneously biasing contacts S$_1$ and S$_2$ and observing interference oscillations at D$_1$ and D$_2$, as shown in Figure \ref{fig:4}c.

The phase shift $\varphi_{\rm{C}}$ induced by the Coulomb interaction is controlled by tuning the channel width by small changes in the upper and lower side-gate voltage $\sim$1 mV. Variations in width of this channel modifies the Coulomb interaction strength via changes in the charge screening strength. Without modifying the Coulomb interaction, the width of the arms should only affect the phase shift $\phi_B$ of the Aharonov-Bohm interference, while the interference visibility is unaffected. This is indeed the case for the single-particle interference, which is barely affected by changes in the side gate voltages $\sim$10 mV \cite{yamamotoElectricalControlSolidstate2012}. By observing the interference oscillations for different side gate voltages, the phase shift $\varphi_{\rm{C}}$ can be adjusted. 
The outcome of two different side-gate conditions are depicted in Figure \ref{fig:4}, where the device is biased from the left, or right side, and then simultaneously from both sides. 
When there is no interaction between electrons (EWPs) or when the phase shift is a multiple of 2$\pi$, the measured currents and interference oscillations should add linearly (Fig. \ref{fig:4}b). 
Interference oscillations that do not correspond to a linear sum for the two-particle injection were observed for certain gate voltages ranges (Fig. \ref{fig:4}a), demonstrating that arbitrary phase shift 
can be induced by the mutual Coulomb interaction of counter-propagating electrons within a coupling length of $\sim 1\mu$m. 
Note that this experiment is performed for the DC transport with the boundary condition set by the measurement configuration and is not a direct demonstration of the two-qubit gate. However, this demonstration of $\varphi_{\rm C}$ shift verifies that the two-qubit gate is possible when two EWPs are incident on the interferometer. 

We believe that the demonstration of the $\varphi_{\rm C}$ shift was made possible by the development of the interferometer operating in the single transmitting mode. In the presence of multiple modes, the effect of the $\varphi_{\rm C}$ shift on interference visibility would be canceled out between modes. This demonstration also provides a pathway to entangle two electron waves through direct Coulomb coupling.

\begin{center}
    \textbf{CONCLUSION}    
\end{center}
Qubit manipulation with enhanced fidelity has been demonstrated using a novel interferometer design that enables control of interference for a single transmitting mode. Furthermore, this device can be operated with ultrafast electronic pulses, and embedded in a loop structure for hosting and manipulating large numbers of time-bin flying qubits. Scalable quantum computing architectures based on this design would consist of multiple parallel loop circuits, interconnected through two-qubit gates.

It is anticipated that millions of flying qubits can be implemented within a hardware architecture that scales as $N^\alpha$, with $\alpha\sim 1/2$ for $\sqrt{N}$ qubits in a single loop, where $N$ is the number of physical qubits and $\alpha$ depends on the number of qubits in a single loop. If one loop can accommodate more than 10,000 qubits, it would be possible to control several million physical qubits using only $\sim$1,000 coaxial lines on a single dilution refrigerator. 
By incorporating delay circuits to induce coupling between non-neighboring distant qubits, the number of the logical qubits can be further increased, or the system could attain the same computational power for much smaller $N$. 
Additionally, we expect the quantum properties of a short EWP to be maintained at significantly higher temperatures than currently possible with DC, allowing for the use of larger cryostats. 
This novel quantum architecture could potentially resolve the scalability challenges faced in the development of quantum computers. 

\begin{figure}[!ht]
    \includegraphics[width=0.45\textwidth]{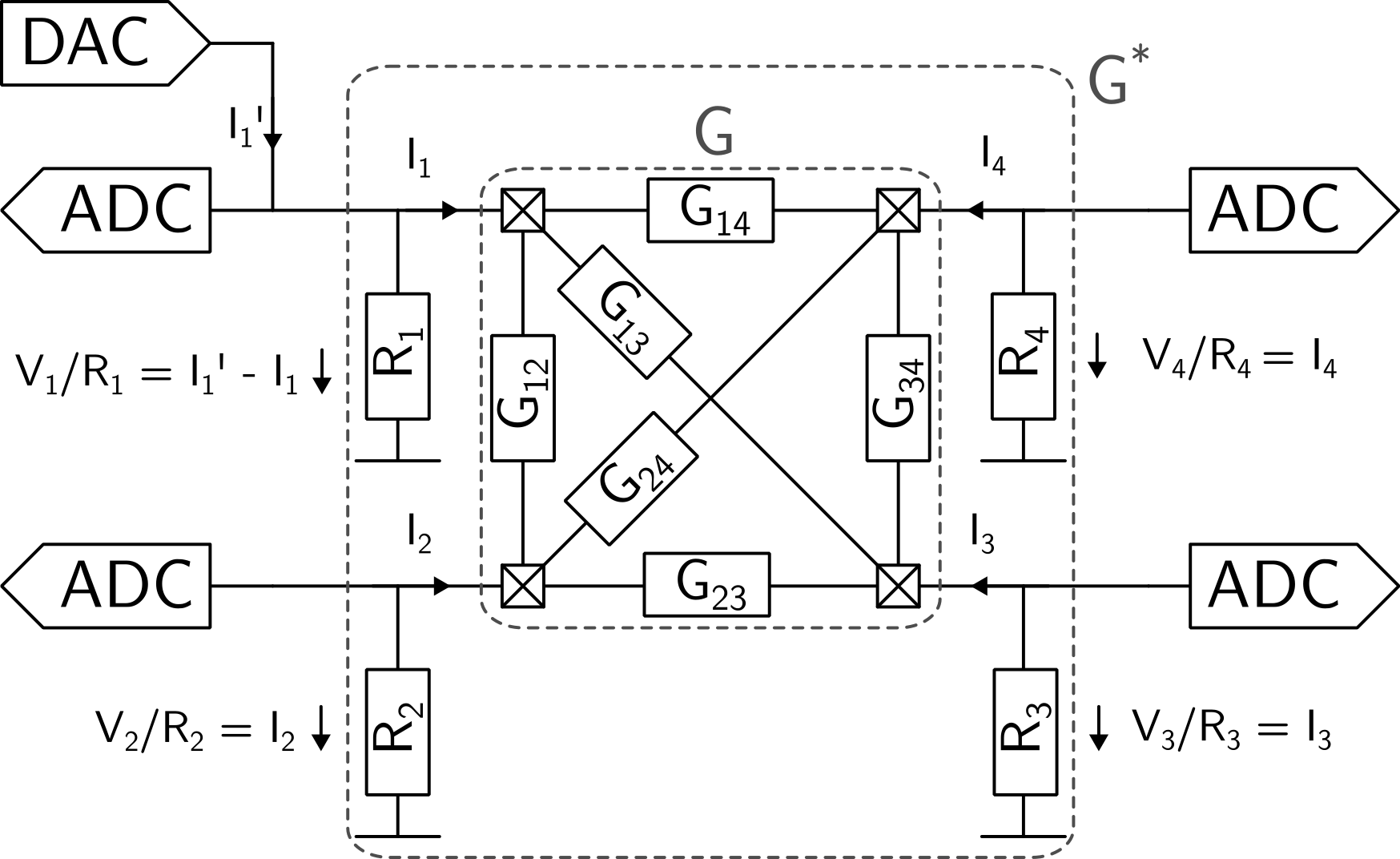}
    \caption{\justifying Conductance model for the four-terminal interferometer. The DAC outputs a constant current $I_1 '$. An ADC measures voltage across current sensing resistors $\{R_1,R_2, R_3, R_4\}$. Lock-in detection is performed with software running on a CPU. The conservation of current at each node is considered to re-construct the conductance matrices $\mathbf G^*$, and $\mathbf G$.}
    \label{supfig:1}
\end{figure}

\begin{center}
    \textbf{SUPPLEMENTARY}
\end{center}
\begin{center}
    \textbf{I. CONDUCTANCE MATRIX}
\end{center}
Measuring the conductance of each path in the electronic interferometer is non-trivial due to the presence of a closed loop, as well as tunnel barriers at the entrance and exit of the four-terminal device. In the regime of a single-channel of quantized conductance, the long paths ($>$1 um) exhibit large conductance fluctuations with small changes in the electrostatic environment, which makes it even more difficult to adjust the conductance of each path. In order to speed up this tuning process, a multi-terminal measurement system was developed to characterize the entire conductance matrix. Supplementary Figure \ref{supfig:1} presents a simplified four-terminal conductance model obtained by considering the conservation of current at each terminal. Terminals 1, 2, 3, and 4 correspond to those in the left upper, lower left, lower right, and upper right contacts in Figure \ref{fig:3}a. Reference current $I_1 '$ generated by a DAC is applied to one terminal, while an ADC acquires the voltage generated at each of the four terminals. Ohmic contacts of the device are grounded at the sample chip carrier through $\{R_1,R_2, R_3, R_4\} = $ 10k$\Omega$ resistors that sense the current flowing through each terminal. 

It is possible to obtain the entire conductance matrix in real-time by applying currents $\{I_1 ', I_2 ', I_3 ', I_4 '\}$ at four distinct frequencies $\{f_1, f_2,f_3,f_4\}$ (typically below 100Hz) to each terminal simultaneously. Dual-phase demodulation is accomplished by measuring the voltage at each terminal with an ADC at a sampling rate of 244kHz, where software-based lock-in detection is performed on a CPU. This setup effectively replaces 16 lock-in amplifiers that would be required to simultaneously detect the amplitude of four distinct frequencies at each of the four terminals.

To obtain the entire conductance matrix $\mathbf G$, first consider a larger conductance matrix defined as $I_i ' = \sum_{j=1}^4 G^*_{ij}V_{j}$, where $I_i '$ is the current applied at terminal $i$, and $V_{j}$ is the voltage measured at terminal $j$. $\mathbf G^*$ is obtained from the inverse of the resistance matrix $\mathbf R^*=(\mathbf G^*)^{-1}$, by applying fixed current $I_i '$ at frequency $f_i$ and measuring the induced voltage $V_{j}(i)$ at each terminal $j$, $R^*_{ji}=V_{j}(i)/I_i ' $. Invoking conservation of current at each of the input nodes, $I_i ' = I_i + V_{i} /R_i$, where $I_i$ is the current flowing into each terminal, and $V_{i} /R_i$ is the current in each resistor grounding the ohmic contacts, these two equations yield the relation $I_i = \sum_{j=1}^4[ \mathbf G^* - \mathbf{1/R_i} ]_{ij} V_{j} $, where $\mathbf G =   G^* -  1/R_i  $, and $\mathbf{ 1/R_i }$ is a diagonal matrix with components $1/R_i$.

For an ideal electronic interferometer where tunnel coupled regions behave as 50-50 beamsplitters, and there is no conduction between adjacent vertical terminals ($G_{12}, G_{34} = 0$, the conductance matrix should have the form 

\begin{equation}\label{eqn:Gideal}
    \begin{split}
    & \mathbf G_\text{ideal} = \frac{2e^2}{h} \times  \\
        &\begin{pmatrix}
        &1 &0 &-\cos^2 (\frac{\phi_B}{2}) &-\sin^2 (\frac{\phi_B}{2})\\
        &0 &1 &-\sin^2 (\frac{\phi_B}{2}) &-\cos^2 (\frac{\phi_B}{2})\\
        &-\cos^2 (\frac{\phi_B}{2}) &-\sin^2 (\frac{\phi_B}{2})  &1  &0 \\
        &-\sin^2 (\frac{\phi_B}{2}) &-\cos^2 (\frac{\phi_B}{2})  &0  &1
        \end{pmatrix},
    \end{split}
\end{equation}

where $\phi_B$ is the relative phase between the upper and the lower channel tuned by the magnetic field. Diagonal elements correspond to the conductance looking into each terminal of the device, such that the row sum is zero, $\sum_{i=1}^{3} {G_{ij}}=1$. 

The measured conductance matrix at a fixed magnetic field, 
\begin{equation}\label{eqn:Gmeas}
    \mathbf G_\text{meas} = \frac{2e^2}{h}
    \begin{pmatrix}
    &1.5 & -&0.4 & -&0.2 & -&0.6\\
    -&0.4 & &1.3 &-&0.5 & -&0.3\\
    -&0.2 & -&0.5  & &1.2  & -&0.3 \\
    -&0.6 & -&0.4  & -&0.3  & &1.3
    \end{pmatrix}
\end{equation}
is indeed fully symmetric, apart from $G_{24} \ne G_{42}$, which is likely due to error in the conversion process.
The deviations from Equation \ref{eqn:Gideal} might arise from multiple loop trajectories enclosing the central Aharanov-Bohm ring. $G_{12} (= G_{21})$ and $G_{34} (= G_{43})$ correspond to the conductance across each tunnel coupled wire. The finite value observed in these elements indicates that there is a leakage of current across each tunnel coupled wire in addition to the interference signal. These effects should be suppressed with the use of electron wave packets (EWPs) rather than DC current. 

\begin{center}
    \textbf{II. ELECTRON WAVE ENTANGLEMENT}
\end{center}
The Coulomb interaction between propagating EWPs is utilized to entangle qubits. The strength of this interaction can be assessed through DC measurements by simultaneously biasing multiple terminals of the interferometer. The resulting $\varphi_{\rm C}$ shift, induced by the Coulomb interaction, is proportional to the length of the interaction region and is influenced by the width of the channels, which modulates the screening effect.

Consider two particles injected from $S_1$ and $S_2$ for the ideal interferometer described in Figure \ref{fig:2}d and \ref{fig:4}a, with the conductance matrix given by Equation \ref{eqn:Gideal}. Two qubits are defined by their presence of a counter-propagating particle in either the upper or lower channels. For simplicity, assume that there is a Coulomb interaction between the two particles only in the Aharonov-Bohm ring when both particles travel through the same arm of the ring. The scattering matrix for the Aharanov-Bohm ring in the two-qubit basis is then given by
\begin{equation}\label{eqn:Uab}
    \mathbf U_\text{AB} = 
    \begin{pmatrix}
    &e^{i(\phi_B-\varphi_C)} & & & & & &\\
    & & &1 & & & &\\
    & & &  & &1  & & \\
    & & &  & &  & &e^{i(-\phi_B-\varphi_C)}
    \end{pmatrix},
\end{equation}
where $\varphi_C$ is the phase shift induced by the mutual Coulomb interaction, and $\phi_B$ is the AB phase shift. The transformation of a beam-splitter in this basis is also given by
\begin{equation} \label{eqn:Ubs}
    \mathbf U_\text{BS} = \frac{1}{2}
    \begin{pmatrix}
    &1 &i &i &-1\\
    &i &1 &-1 &i\\
    &i &-1  &1  &i \\
    &-1 &i  &i  &1
    \end{pmatrix}.
\end{equation}
For the case of an initial state where both particles traverse the upper path, $\lvert 0_1 1_2 \rangle$, the final state is fully entangled for $\varphi_C=\pi/2+n\pi$, whereas there is no entanglement for $\varphi_C=n\pi$, where $n$ is integer. For example, the final state for $\varphi_C=\pi/2$ is 
\begin{equation}\label{eqn:EntangledState}
\begin{split}\psi_f=\frac{1}{2}\ket{0}_1[(-\sin \phi_B \ket{0}_2+(\cos \phi_B +i)\ket{1}_2]\\+\frac{1}{2}\ket{1}_1 [(-\cos \phi_B +i)\ket{0}_2-\sin \phi_B \ket{1}_2].
\end{split}
\end{equation}
The current at $D_1$ and $D_2$ are independent of $\phi_B$ due to the cancellations between the first and second terms of Equation \ref{eqn:EntangledState}, causing the interference signal to vanish.

The strength of the Coulomb interaction, $\varphi_C$, also depends on the width of the channel and is sensitive to the side gate voltages. Our results show that the change in $\varphi_C$ is much larger than that of $\phi_B$ for the applied source voltages of approximately 10 $\mu$V when the side gate voltage is swept.

This interference can also be interpreted as a mutual quantum gate voltage for electron waves. The electron wave injected from S$_1$ (S$_2$) is modulated by the gate voltage applied to S$_2$ (S$_1$) in a quantum manner, representing two-particle interference, independent of the voltages at S$_1$ and S$_2$.

\begin{figure}[!ht]
    \includegraphics[width=0.5\textwidth]{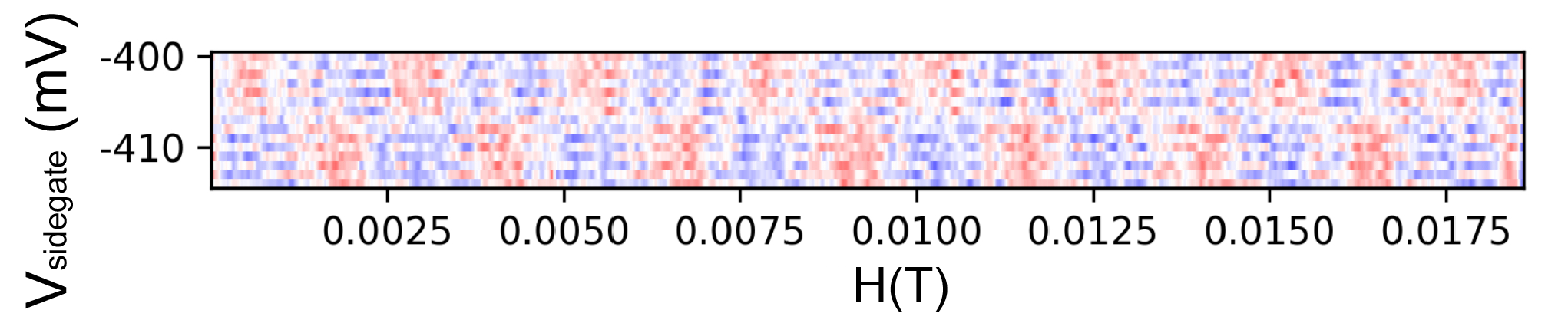}
    \caption{\justifying Measured current (arbitrary units) for in-phase oscillations in a two-electron experiment as a function of the lower side gate voltage (see Figure \ref{fig:4}c).}
    \label{supfig:2}
\end{figure}

\begin{center}
    \textbf{III. PHASE RIGIDITY}
\end{center}
Figure \ref{supfig:2} shows Aharonov-Bohm (AB) oscillations from a two-electron experiment as a function of one of the side gate voltages, $V_{\rm sidegate}$. The interaction is controlled by adjusting the channel width via the side gate voltage. In this experiment, the total current oscillates as a function of the magnetic field due to the interference of multiple paths enclosing the ring. This creates a standing wave within the ring, resulting in the rigidity of the observed AB oscillation phase \cite{Takada2015}. For single-mode interference, only abrupt phase shifts by $\pi$ are allowed, while smooth phase shifts occur when multiple modes and inter-mode scattering are present. The observation of only abrupt jumps in the phase shift confirms the absence of inter-mode scattering and indicates the contribution of a single mode.

\bibliography{references}

\end{document}